\documentclass[onecolumn,floatfix,ams,nofootinbib]{revtex4}
\pdfoutput=1
\usepackage[dvips]{epsfig}
\usepackage[english]{babel}
\usepackage[utf8]{inputenc}
\usepackage{bbm}
\usepackage{verbatim}
\usepackage{array}
\usepackage{bm} 
\usepackage{amsmath}
\usepackage{yfonts}
\usepackage{amsthm}
\usepackage{amsmath,amscd}
\usepackage{pst-plot}
\usepackage{slashed} 
\usepackage{tikz-cd}
\usepackage{amsfonts}
\usepackage{graphicx}
\usepackage{amssymb}
\newtheorem{definition}{Definition} 
\usepackage{hyperref}
\usepackage{cases}
\usepackage{indentfirst} 
\usepackage[titletoc]{appendix}

\usepackage{subeqnarray}
\usepackage{setspace}
\usepackage{indentfirst} %primeiro paragrafo com espaço

\usepackage{gensymb}%simbolo º
\usepackage{xcolor}%Pacote de mudança de cores nos caracteres 
\usepackage{calrsfs}%Fonte do operador Dilatação

\usepackage{wasysym}
\usepackage[all,cmtip]{xy}

    \usepackage{amsmath,amsfonts,amssymb,amsthm,mathrsfs,bbm,braket}

    \newtheorem{prop}{Proposition}[section]
    \newtheorem{theo}{Theorem}[section]
    \newtheorem{lemma}{Lemma}[section]

    \numberwithin{equation}{section}

\usepackage{bbold}

\IfFileExists{ulem.sty}{\usepackage[normalem]{ulem}}{\newcommand{\sout}[1]{#1}}   % strike-through; plain red if ulem is missing
\newif\ifshowchanges
\showchangestrue
\ifshowchanges
  \newcommand{\del}[1]{{\color{red}\sout{#1}}}      % removed plain text (inline)
  \newcommand{\delm}[1]{{\color{red}#1}}            % removed math / text with cites, footnotes
  \newenvironment{newtext}{\color{blue}}{}          % added block
  \newenvironment{oldtext}{\color{red}}{}           % removed block
\else
  
  \newcommand{\del}[1]{}
  \newcommand{\delm}[1]{}
  
\fi
\theoremstyle{definition}

\theoremstyle{plain}

\newcommand{\GLp}[1]{\mathrm{GL}^{+}(#1,\mathbb{R})}
\newcommand{\GLpt}[1]{\widetilde{\mathrm{GL}}^{+}(#1,\mathbb{R})}
\newcommand{\HM}{T\mathcal{M}\oplus_W T^{*}\mathcal{M}}

\begin{document}

\title{Notes on obstructions in the hyperbolic Clifford algebra bundle structure} 

\author{J. M. Hoff da Silva} 
\email{julio.hoff@unesp.br}
\affiliation{Departamento de F\'isica, Universidade
	Estadual Paulista, UNESP, Av. Dr. Ariberto Pereira da Cunha, 333, Guaratinguet\'a, SP,
	Brazil}

\author{E. Notte-Cuello} 
\email{enotte@userena.cl}
\affiliation{Departamento de Matem\'aticas, Universidad
	de La Serena, Av. J. Cisternas 1200, Chile}

\begin{abstract}
Starting from a general analysis of obstruction classes, we develop the investigation of obstructions associated with the bundle structure of the hyperbolic Clifford algebra. By taking into account particularities arising from the Whitney sum, it is shown that, unlike classical tangent bundle cases, the hyperbolic frame bundle of an oriented manifold admits lifting without any topological obstruction, that is, it admits a structure of the connected double covering of $\mathrm{GL}^{+}(2m,\mathbb{R})$, and the hyperbolic tangent bundle carries a $Spin^{e}_{m,m}$ structure (with $m=\dim\mathcal{M}$), whatever the value of $w_2(\mathcal{M})$. These conclusions are known from the work of Karoubi and Gualtieri and the point of this note is that they follow from the Whitney product formula alone, or from the additivity formula of Whitney sums with an explicit covering group, without recourse to Karoubi's criterion for indefinite signature. We also describe the spinor bundles involved, which are bundles of differential forms and not Dirac spinor bundles and show that orientability is needed for the covering of the full linear group.
\end{abstract}

\maketitle

\section{Introduction}

The question of whether a smooth manifold admits a spinor structure is one of the most fundamental problems at the interface of differential geometry, algebraic topology, and mathematical physics. In the classical setting, a spacetime manifold $\mathcal{M}$, understood as a 4-dimensional manifold equipped with a Lorentzian metric of signature $(1,3)$, can support globally defined spinor fields only if it is endowed with additional structure beyond its metric: the so-called spinor structure. The systematic investigation of this requirement was initiated by Geroch in his landmark two-part study ~\cite{Geroch1, Geroch2}, where he proved that a necessary and sufficient condition for a noncompact spacetime to admit a spinor structure is that it carry a global field of orthonormal tetrads. Equivalently, this condition is captured by the vanishing of the second Stiefel–Whitney class $w_2(\mathcal{M}) \in \check{H}^2(\mathcal{\mathcal{M}},\mathbb{Z}_2)$, an obstruction class whose triviality is required for the transition functions of the frame bundle $P_{SO^e_{1,3}}(M)$ to lift to the double cover $Spin^e_{1,3}$. The structure of such obstructions, and the machinery of lifting structure groups via central homomorphisms, was subsequently developed in greater generality by Greub and Petry ~\cite{GP}, providing the algebraic-topological framework on which the present work is based.

In parallel with these foundational developments, the Clifford bundle formalism has emerged as a powerful and geometrically transparent alternative to tensor calculus for the study of gravitational fields and spinor fields on curved spacetimes. As described in ~\cite{RS0, NR}, the Clifford bundle $Cl({T}^{*}\mathcal{M}, g)$ over a Lorentzian manifold provides a natural arena for reformulating Einstein’s gravitational theory, Yang–Mills-type gravitational Lagrangians, and the Dirac equation in a unified geometric language. Within this formalism, the gravitational field is modeled not merely as a metric structure but as a full geometric entity intrinsic to the bundle structure of spacetime.

A significant generalization of this framework arises upon replacing the standard Clifford algebra $Cl(V, g)$, built on the vector space $V$ alone,  by the Hyperbolic Clifford algebra $Cl(H_V)$, which is naturally associated with the hyperbolic space $H_V=V\oplus V^{*}$ endowed with the canonical neutral bilinear form $\langle x, y\rangle=x^{*}(y_*)+y^{*}(x_*)$.  This algebra, first introduced in ~\cite{RSV} and developed extensively in ~\cite{RS}, simultaneously accommodates both multivectors and multiforms within a single unified algebraic structure. Its fundamental isomorphism $Cl(H_V)\simeq\operatorname*{End}\bigwedge V$ earns it the designation of {\it mother algebra} of the vector space $V$, in the sense that it contains, as subalgebras, the standard Clifford algebras $Cl(V, b)$ and $Cl(V, -b)$ for any non-degenerate symmetric bilinear form $b$ on $V$. Beyond its algebraic elegance, this structure is physically motivated: it has been shown ~\cite{RS} that the space $\bigwedge V^{*}$  constitutes a minimal left ideal of $Cl(H_V)$, whose elements are representatives of Witten superfields, making the hyperbolic Clifford algebra the natural algebraic setting for the study of superfields in theoretical physics.

The globalization of this algebraic structure to a smooth $m$-dimensional manifold $\mathcal{M}$ was carried out in ~\cite{nc}, where the hyperbolic tangent bundle $T\mathcal{M} \oplus T^{*}\mathcal{M}$ was shown to be associated with the principal hyperbolic frame bundle $F_H(\mathcal{M}) = \cup_{x \in \mathcal{M}} (F_x\mathcal{M} \oplus F^*_x\mathcal{M})$ with structure group $\mathrm{GL}(2m, \mathbb{R})$. The introduction of a metric field and the subsequent reduction of this frame bundle to the oriented hyperbolic orthonormal frame bundle $P_{SO^{e}_{m,m}}(\mathcal{M})$ then opens the door to defining hyperbolic spinor structures, by analogy with Geroch’s classical construction  ~\cite{Geroch1, Geroch2}, via the covering group $Spin^{e}_{m,m}$. However, as remarked in ~\cite{nc}, the precise topological obstructions to the existence of such hyperbolic structures, and whether they are more or less restrictive than those in the classical case, were left as an open problem for future investigation. 

Karoubi \cite{kar} gave a criterion for an oriented vector bundle $E$ with structure group $SO(p,q)$ to admit a $Spin(p,q)$ structure: with $E=E^{+}\oplus E^{-}$ the splitting into positive and negative definite subbundles given by a reduction to the maximal compact subgroup, the condition is $w_2(E^{+})=w_2(E^{-})$. This criterion is reproduced in Gualtieri's thesis \cite{gualth}. In the thesis it is applied to $T\oplus T^{*}$: the bundles $E^{\pm}$ project isomorphically onto the tangent bundle, so the condition holds, and it is then proved that $T\oplus T^{*}$ always admits a $Spin(m,m)$ structure. Moreover, the same thesis also exhibits an explicit splitting of the restriction of the spin covering to $\mathrm{GL}(V)\subset SO(V\oplus V^{*})$ and identify the resulting spinor bundles with $\wedge^{\bullet}T^{*}\otimes|\det T|^{1/2}$, in two versions that differ by a real line bundle. The existence of the relevant spinor structures is thus known, and the spinor representation on $\wedge^\bullet T^*$ is a standard ingredient of generalized geometry \cite{gualt}. The results below must be read in this light.

The present paper takes up this problem in generality. We systematically investigate the obstruction theory for the hyperbolic Clifford algebra bundle structure, working within the framework of Čech cohomology and the theory of lifting structure groups. Our approach begins with a general analysis of obstruction classes for finite collections of orientable principal bundles over a common base manifold $\mathcal{M}$, following the formalism of Greub and Petry ~\cite{GP}, and derives an additivity formula for the obstruction classes associated with their Whitney sum.  Specifically, if $\{\bar{P}_{i}\}$ is a collection of orientable principal bundles with structure groups $G_{i}$ and central homomorphisms 
$\rho_i:\mathbb{G}_i\rightarrow G_i$ with discrete abelian kernels $K_i $ in the center of $\mathbb{G}_i$, we show that, once a surjective homomorphism $\varphi:K_1\times\cdots\times K_n\rightarrow K$ is chosen to specify the covering group of the product, the obstruction class associated with the Whitney sum satisfies the additivity relation $\kappa[(\oplus_W)_{i=1}^n P_i]=\sum_{i=1}^n\kappa^i(P_i)\in \check{H}^2(\mathcal{M},\mathbb{Z}_2)$ (for $K_i=\mathbb{Z}_2$ and $\varphi$ the sum). We then specialize this framework to the hyperbolic frame bundle $F_H(\mathcal{M})$, viewed as the Whitney sum $T\mathcal{M} \oplus_W T^{*}\mathcal{M}$. The central result of this paper (Theorem~\ref{thm:main}) is that, unlike the classical tangent bundle $T\mathcal{M}$, whose spin structure requires the non-trivial topological condition $w_2(\mathcal{M}) = 0$, the hyperbolic frame bundle $F_H(\mathcal{M})$ of an oriented manifold admits a lifting of its structure group to the connected double covering $\widetilde{\mathrm{GL}}^{+}(2\dim\mathcal{M},\mathbb{R})$ of $\mathrm{GL}^{+}(2\dim\mathcal{M},\mathbb{R})$ without any topological obstruction. This absence of obstruction follows from a cancellation mechanism intrinsic to the Whitney sum: since $T\mathcal{M}$ is isomorphic to $T^{*}\mathcal{M}$ as vector bundles over any paracompact manifold (by means of a Riemannian metric), the two summands contribute equal obstruction classes in $\check{H}^2(\mathcal{M},\mathbb{Z}_2)$, and their sum vanishes modulo 2.

These results answer the question left open in \cite{nc} for oriented manifolds. We remark that the obstruction-vanishing mechanism established here is directly inherited from two elementary ingredients: the multiplicative behavior of Stiefel–Whitney classes under Whitney sums, and the isomorphism between $T\mathcal{M}$ and $T^{*}\mathcal{M}$ valid on any paracompact manifold. In this sense, part of our contribution consists in generalizing the obstruction formalism presented in Ref. \cite{GP} to arbitrary (finite) Whitney sums of orientable principal bundles with independent central homomorphisms and possibly distinct kernels, with the target covering group made explicit, a generalization that allows classical results to be recovered more or less straightforwardly. More importantly, applying the resulting analysis to the question raised in Ref. \cite{nc} shows the complete absence of obstructions for the hyperbolic frame bundle of an oriented manifold. We observe that the vanishing of the obstruction is a consequence of the Whitney product formula, and the existence of a $Spin(m,m)$ structure on $T\mathcal{M}\oplus T^{*}\mathcal{M}$ is known \cite{kar,gualth} and we claim no priority for these conclusions. What we do claim is that they are obtained here by considerably more elementary means than an appeal to Karoubi's criterion for indefinite signature: the covering of $\mathrm{GL}^{+}(2m,\mathbb{R})$ is handled by the Whitney product formula alone (or, equivalently, by the triviality of that covering over $\mathrm{GL}^{+}(m,\mathbb{R})$), and the $Spin^{e}_{m,m}$ statement follows from Proposition~\ref{prop:spin0} in a few lines, because the reduction of the hyperbolic frame bundle to its maximal compact subgroup is the fibre product of two copies of the same bundle of orthonormal frames of $\mathcal{M}$ (Section~\ref{sec:ortho}). 

The paper is organized as follows. In Section~\ref{sec:prelim} we fix the notation and recall the hyperbolic frame bundle and the covering groups. In Section \ref{sec:obstr}, we develop the general obstruction theory for Whitney sums of principal bundles, establishing the additivity formula for obstruction classes with an explicit target group. We then apply this framework to the hyperbolic frame bundle, establish the absence of obstructions to the hyperbolic spin structure and discuss the classification of non-equivalent spinor structures in terms of the fundamental group of the base manifold. Section~\ref{sec:ortho} treats the passage to orthonormal frames and the $Spin^{e}_{m,m}$ structure, the associated spinor bundles. In Section~\ref{sec:concl} we conclude.

\section{Preliminaries: the hyperbolic space and covering groups}\label{sec:prelim}

In this section we fix the notation and recall the geometric objects and the covering groups that enter the analysis. We start requiring only $m=\dim\mathcal{M}\geq 2$ but the fundamental-group arguments, Lemma \ref{lem:pi1} for instance, additionally require $m\geq 3$.

Starting with the hyperbolic space revision, let $V$ be a real vector space of dimension $m$ and $H_V=V\oplus V^{*}$ the hyperbolic space of the Introduction, whose elements we write $x=x_{*}+x^{*}$ with $x_{*}\in V$ and $x^{*}\in V^{*}$, endowed with the canonical neutral bilinear form $\langle x,y\rangle=x^{*}(y_{*})+y^{*}(x_{*})$. Both $V$ and $V^{*}$ are maximal isotropic subspaces, and the signature is $(m,m)$. Given a positive definite inner product $g$ on $V$, put
\begin{equation}
E^{\pm}=\{X\pm g(X):X\in V\},\qquad H_V=E^{+}\oplus E^{-},\label{eq:Epm}
\end{equation}
where $g(X)\in V^{*}$ denotes $g(X,\cdot)$. Then, naturally, $\langle X\pm g(X),X\pm g(X)\rangle=\pm2g(X,X)$ and $\langle X+g(X),Y-g(Y)\rangle=0$, so $E^{+}$ is positive definite, $E^{-}$ is negative definite, the two are orthogonal, and the projection $H_V\rightarrow V$ restricts to isometries $(E^{\pm},\pm\frac12\langle\,,\rangle)\simeq(V,g)$. This splitting is the one we shall use in the reduction to the maximal compact subgroup. The Clifford algebra $Cl(H_V)$ acts on $\bigwedge V^{*}$ by $(x_{*}+x^{*})\cdot\varphi=i_{x_{*}}\varphi+x^{*}\wedge\varphi$, whose square is $x^{*}(x_{*})=\frac12\langle x,x\rangle$ (the normalization of the Clifford relation is immaterial in what follows). This gives the isomorphism $Cl(H_V)\simeq\operatorname*{End}\bigwedge V^{*}$ of the mother algebra \cite{RSV,RS}. Since $E^{+}\perp E^{-}$ we also have $Cl(H_V)\simeq Cl(E^{+})\hat{\otimes}Cl(E^{-})\simeq Cl(V,g)\hat{\otimes}Cl(V,-g)$ (graded tensor product), which exhibits the standard Clifford algebras of \cite{RS0,NR} inside $Cl(H_V)$.

The hyperbolic tangent bundle is $\HM$, with the form above on each fibre. Throughout the main text $\mathcal{M}$ is assumed to be oriented (this assumption is further discussed along the text) and we denote by $F^{+}(\mathcal{M})$ its principal $\GLp{m}$-bundle of oriented frames. An oriented frame $(e_i)$ of $T\mathcal{M}$ determines the dual coframe $(e^i)$ of $T^{*}\mathcal{M}$ and the pair $(e_i,e^i)$ is a frame of $\HM$. Let
\begin{equation}
\iota:\GLp{m}\rightarrow\GLp{2m},\qquad \iota(A)=\mathrm{diag}\big(A,(A^{-1})^{T}\big),\label{eq:iota}
\end{equation}
be a continuous homomorphism. Then $\HM\simeq F^{+}(\mathcal{M})\times_{\iota}\mathbb{R}^{2m}$ is oriented and the hyperbolic frame bundle $F_H(\mathcal{M})$ of the Introduction is the principal $\GLp{2m}$-bundle of oriented frames of $\HM$, i. e., $F_H(\mathcal{M})=F^{+}(\mathcal{M})\times_{\iota}\GLp{2m}$. If one regards $T\mathcal{M}$ and $T^{*}\mathcal{M}$ as independent oriented bundles, the same bundle is obtained from the Whitney sum, a principal $\GLp{m}\times\GLp{m}$-bundle, by extension along the block-diagonal inclusion
\begin{equation}
j:\GLp{m}\times\GLp{m}\rightarrow\GLp{2m},\qquad j(A,B)=\mathrm{diag}(A,B).\label{eq:j}
\end{equation}
Note that $\iota=j\circ(\mathrm{id},\tau)\circ\Delta$, with $\Delta$ the diagonal map and $\tau(A)=(A^{-1})^{T}$.

Turning to covering groups, the group $\GLp{k}$ deformation retracts onto $SO(k)$ (via polar decomposition), so that $\pi_1(\GLp{k})\simeq\pi_1(SO(k))$ which is $\mathbb{Z}$ for $k=2$ and $\mathbb{Z}_2$ for $k\geq3$. Besides, since $\mathrm{GL}(k,\mathbb{R})$ is disconnected, the double cover is not uniquely determined and we always work on the identity component. For $k\geq2$ we denote by $p:\GLpt{k}\rightarrow\GLp{k}$ the connected double covering. It is unique, for $k\geq 3$ it is the universal covering, and its restriction to $SO(k)$ is simply the double covering $Spin(k)\rightarrow SO(k)$. The group of linear automorphisms of $H_V$ preserving $\langle\,,\rangle$ and the canonical orientation is $SO(m,m)$, which has two connected components. We write $SO^{e}_{m,m}$ for its identity component and $Spin^{e}_{m,m}$ for the identity component of $Spin(m,m)\subset Cl(H_V)$, a connected double covering of $SO^{e}_{m,m}$. The maximal compact subgroup of $SO^{e}_{m,m}$ is $SO(m)\times SO(m)$, acting on $E^{+}\oplus E^{-}$, and the corresponding maximal compact subgroup of $Spin^{e}_{m,m}$ is $Spin(m)\times_{\mathbb{Z}_2}Spin(m):=(Spin(m)\times Spin(m))/\{(1,1),(-1,-1)\}$. The homomorphism $\iota$ of (\ref{eq:iota}) takes values in $SO^{e}_{m,m}$, since $\GLp{m}$ is connected and on the maximal compact subgroup it maps $A\in SO(m)$ to $(A,A)\in SO(m)\times SO(m)$, because $(A^{-1})^{T}=A$ for orthogonal $A$. These specifications allows the following lemma.

\begin{lemma}\label{lem:pi1}
For $m\geq3$:
\begin{itemize}
\item[(i)] the block-diagonal inclusion $j$ of (\ref{eq:j}) induces $j_{*}:\mathbb{Z}_2\times\mathbb{Z}_2\rightarrow\mathbb{Z}_2$, $(x,y)\mapsto x+y$;
\item[(ii)] the automorphism $\tau(A)=(A^{-1})^{T}$ induces the identity on $\pi_1(\GLp{m})\simeq\mathbb{Z}_2$;
\item[(iii)] consequently the homomorphism $\iota$ of (\ref{eq:iota}) induces the zero map $\iota_{*}:\pi_1(\GLp{m})\rightarrow\pi_1(\GLp{2m})$.
\end{itemize}
\end{lemma}

\begin{proof}
By polar decomposition it suffices to work with $SO(m)$ and $SO(2m)$. (i) Write $j(A,B)=\epsilon(A)\epsilon'(B)$, where $\epsilon(A)=\mathrm{diag}(A,1)$ and $\epsilon'(B)=\mathrm{diag}(1,B)$ are commuting homomorphisms $SO(m)\rightarrow SO(2m)$. In a topological group, the pointwise product $\alpha\cdot \beta$ of two loops at $e$ is homotopic to their concatenation $\alpha *\beta$, hence $[\alpha\cdot\beta]=[\alpha]+ [\beta]$. Therefore $j_{*}(x,y)=\epsilon_{*}(x)+\epsilon'_{*}(y)$. The inclusion $SO(k)\rightarrow SO(k+1)$ induces an isomorphism on $\pi_1$ for $k\geq3$, hence $\epsilon_{*}$ is an isomorphism $\mathbb{Z}_2\rightarrow\mathbb{Z}_2$, the same holding for $\epsilon'$ \cite{ao}. Since $\pi_1(SO(m))$ and $\pi_1(SO(2m))$ have order $2$, $\epsilon$ and $\epsilon'$ are the unique isomorphism $\mathbb{Z}_2\rightarrow \mathbb{Z}_2$, that is, the identity. Thus $j_{*}(x,y)=x+y$. (ii) On $SO(m)$ one has $\tau(A)=A$, so $\tau_{*}=\mathrm{id}$. (iii) With $\iota=j\circ(\mathrm{id},\tau)\circ\Delta$ we get $\iota_{*}(x)=j_{*}(x,\tau_{*}x)=x+x=0$.  
\end{proof}

\section{Obstruction classes related to the bundle structure of the Hyperbolic Clifford algebra}\label{sec:obstr}

We shall initially perform the analysis on a general basis, in a manner akin to  Ref.\cite{GP}, after which we particularize it to the cases at hand. Consider $\{\bar{P}_{i}\}$ to be a set of $n \in \mathbb{N}^{*}$ orientable principal bundles over the same manifold $\mathcal{M}$, namely $\bar{P}_{i}=(\bar{P}_{i}, \mathcal{M}, \pi_{i}, G_{i})$. Moreover, let $\{\rho_{i}\}$ be a set of central homomorphisms from $\mathbb{G}_{i}$ to $G_{i}$; that is, $\ker \rho_{i}=K_{i}$, where $K_{i}$ is an abelian discrete group belonging to the center of $\mathbb{G}_{i}$ for each $i=1, \dots, n$.

\begin{definition}
	A $\mathbb{G}_i$-structure on $\bar{P}_i$ is defined, for each $i=1,\cdots,n$, as a principal bundle $\bar{P}_{\mathbb{G}_i}=(\bar{P}_{\mathbb{G}_i},\mathcal{M},\pi_{\mathbb{G}_i},\mathbb{G}_i)$ together with an equivariant bundle map $\eta_i:\bar{P}_{\mathbb{G}_i}\rightarrow \bar{P}_i$, such that the relation $\eta_i(l\cdot \gamma)=\eta_i(l)\cdot \rho_i(\gamma)$ holds for any $l\in \bar{P}_{\mathbb{G}_i}$ and $\gamma\in \mathbb{G}_i$.  	
\end{definition} 

In the above definition, $\mathbb{G}_i$ is the covering group and $G_i$ is the structure group of $\bar{P}_i$. They are different groups, related by the central homomorphism $\rho_i:\mathbb{G}_i\rightarrow G_i$, so that $\gamma\in\mathbb{G}_i$ whereas $\rho_i(\gamma)\in G_i$. Throughout this work, we assume that $\mathcal{M}$ is orientable and admits a good cover. Naturally, let $\{\sigma^i_\alpha\}$ be the local trivializations for $P_i$ subordinated to the same open cover $\{U_\alpha\}$ ($\alpha\in I$) of $\mathcal{M}$. For $x\in U_\alpha\cap U_\beta\subset \mathcal{M}$, we define the transition functions $s_{\alpha\beta}^i(x)=\sigma^i_{\alpha}(x)\circ(\sigma^i_{\beta})^{-1}(x)$, which satisfy the consistency relations $s_{\alpha\beta}^i(x)s_{\beta\lambda}^i(x)=s_{\alpha\lambda}^i(x)$ for $x\in U_\alpha\cap U_\beta\cap U_\lambda$. With this construction, standard results on $\mathbb{G}_i$-structures for $\bar{P}_i$ apply. In particular, a given bundle $\bar{P}_i$ admits a $\mathbb{G}_i$-structure if, and only if, there exist continuous maps $\gamma^i_{\alpha\beta}:U_\alpha\cap U_\beta\rightarrow \mathbb{G}_i$ such that the consistency condition holds for the $\{\gamma^i_{\alpha\beta}\}$ and $\rho^i(\gamma_{\alpha\beta}^i)=s_{\alpha\beta}^i$. When this condition is met, the transition maps $s_{\alpha\beta}^i: U_\alpha\cap U_\beta\rightarrow G_i$ are said to lift to the continuous maps $\gamma^i_{\alpha\beta}: U_\alpha\cap U_\beta\rightarrow\mathbb{G}_i$. Now, for $x\in U_\alpha\cap U_\beta\cap U_\lambda$, we define:
\begin{equation}
q^i_{\alpha\beta\lambda}=\gamma^i_{\beta\lambda}(\gamma^i_{\alpha\lambda})^{-1}\gamma^i_{\alpha\beta}. 
\end{equation} It is straightforward to verify that $\rho^i(q^i_{\alpha\beta\lambda})=s^i_{\beta\lambda}s^i_{\lambda\alpha} s^i_{\alpha\beta}=e^i$, where $e^i$ denotes the identity element of $G_i$. Consequently, $q^i_{\alpha\beta\lambda} \in K_i$, which defines a $2$-cochain on $\mathcal{M}$ with values in $K_i$. Furthermore, let us denote such a cochain by $q^i(\alpha,\beta,\lambda)$ and the coboundary operator by $\delta$. Taking into account that $K_i$ is abelian and lies in the center of $\mathbb{G}_i$, it can be directly verified that
\begin{equation}
\delta q^i(\alpha,\beta,\lambda,\mu)=q^i(\beta,\lambda,\mu)(q^i(\alpha,\lambda,\mu))^{-1}q^i(\alpha,\beta,\mu)(q^i(\alpha,\beta,\lambda))^{-1}=e^i
\end{equation} and $q^i(\alpha,\beta,\lambda)$ is a cocycle. The existence of $q^i$ determines an element $\kappa^i \in \check{H}^2(\mathcal{M}, K_i)$, which acts as an obstruction class. In fact, a given bundle $\bar{P}_i$ admits a $\mathbb{G}_i$-structure if, and only if, $\kappa^i$ is trivial (or null) ~\cite{td}. 

Moving forward, let $(\oplus_W)_{i=1}^n$ denote the Whitney sum ~\cite{ws} of the vector bundles $\{P_i\}$ associated with $\{\bar{P}_i\}$ for each $i=1, \dots, n$. Thus, the Whitney sum bundle is defined as
\begin{equation}
(\oplus_W)_{i=1}^n P_i=\{(p_1,\cdots,p_n) \in P_1\times\cdots\times P_n:\pi_1(p_1)=\cdots=\pi_n(p_n)\},
\end{equation} with projections naturally given by 
\begin{eqnarray}
\Pi_W&:&\left. (\oplus_W)_{i=1}^n P_i\rightarrow \mathcal{M}\right.\nonumber\\&&
\left. (p_1,\cdots,p_n)\mapsto \Pi_W(p_1,\cdots,p_n)=\pi_1(p_1)=\cdots=\pi_n(p_n).\right.
\end{eqnarray} We emphasize that the results discussed above concerning obstructions also apply to the set $\{P_i\}$, as they pertain to the lifting of structure groups. Furthermore, the trivializations for $(\oplus_W)_{i=1}^n P_i$ are given by the Cartesian product of the individual trivializations, $\{\sigma^1_\alpha \times \dots \times \sigma^n_\alpha\} \equiv \{(\times)_{i=1}^n \sigma_\alpha\}$. The composition of the central homomorphisms is then given by
\begin{eqnarray}
(\times)_{i=1}^n\rho_i:(\times)_{i=1}^n\mathbb{G}_i\rightarrow (\times)_{i=1}^n G_i,  
\end{eqnarray} such that $\ker((\times)_{i=1}^n\rho_i)=(\times)_{i=1}^n K_i$. Due to the natural gluing of fibers in the Whitney sum, the transition functions for $(\oplus_W)_{i=1}^n P_i$ are given by
\begin{equation}
g_{\alpha\beta}=(\oplus_W)_{i=1}^n s^i_{\alpha\beta}=\text{diag}(s^1_{\alpha\beta}, \cdots, s^n_{\alpha\beta}),
\end{equation} whose explicit form is inherited from the embedding of $(\times)_{i=1}^n G_i$. In fact, for $x\in U_\alpha\cap U_\beta\cap U_\lambda$
\begin{eqnarray}
g_{\alpha\beta}g_{\beta\lambda}=\text{diag}(s^1_{\alpha\beta}s^1_{\beta\lambda}, \cdots, s^n_{\alpha\beta}s^n_{\beta\lambda})=\text{diag}(s^1_{\alpha\lambda}, \cdots, s^n_{\alpha\lambda})=(\oplus_W)_{i=1}^n s^i_{\alpha\lambda}=g_{\alpha\lambda}.  
\end{eqnarray} 

Recall that we have assumed all bundles in $\{P_i\}$ to be orientable. This avoids subtleties arising from cup products between the first Stiefel-Whitney classes when computing the obstruction classes. Hence, a more direct approach applies. In this vein, let $C^p(U \subset \mathcal{M}, K)$ be the multiplicative Čech cochain group. Therefore, if $f_i \in C^p(U, K_i)$, then
\begin{eqnarray}
\Big(\bigoplus_{i=1}^n f_i\Big)(i_0\cdots i_p)=\bigoplus_{i=1}^n f_i(i_0\cdots i_p)\in \bigoplus_{i=1}^n C^p(U,K_i),
\end{eqnarray} where $\bigoplus_{i=1}^n$ stands for the usual direct sum. Additionally, defining $\gamma_{\alpha\beta}:U_\alpha\cap U_\beta\rightarrow (\times)_{i=1}^n\mathbb{G}_i$ we arrive at 
\begin{eqnarray}
\rho\gamma_{\alpha\beta}:=(\times)_{i=1}^n\rho_i[(\oplus_W)_{i=1}^n \gamma^i_{\alpha\beta}]=(\oplus_W)_{i=1}^n \rho_i \gamma^i_{\alpha\beta}=(\oplus_W)_{i=1}^n s^i_{\alpha\beta}=g_{\alpha\beta} 
\end{eqnarray} and therefore $\gamma_{\alpha\beta}$ are indeed the lifts of $g_{\alpha\beta}$. Notice that
\begin{eqnarray}
\gamma_{\beta\lambda}\gamma^{-1}_{\alpha\lambda}\gamma_{\alpha\beta}=[(\oplus_W)_{i=1}^n \gamma_{\beta\lambda}^i][(\oplus_W)_{i=1}^n \gamma^i_{\alpha\lambda}]^{-1}[(\oplus_W)_{i=1}^n \gamma_{\alpha\beta}^i]=(\oplus_W)_{i=1}^n \gamma^i_{\beta\lambda}(\gamma^{i}_{\alpha\lambda})^{-1}\gamma^i_{\alpha\beta},\label{exp}
\end{eqnarray} since $[(\oplus_W)_{i=1}^n \gamma^i_{\alpha\lambda}]^{-1}=[(\oplus_W)_{i=1}^n (\gamma^i_{\alpha\lambda})^{-1}]$. Expression (\ref{exp}) contains nothing but the individual cocycles associated with the bundles forming the Whitney sum. Therefore, the obstruction class associated with the total sum can be read off from the $n$-tuple $(q^1,\ldots,q^n)$, which is a cocycle with values in $K_1\times\cdots\times K_n$ for the product homomorphism $(\times)_{i=1}^n\rho_i$. In order to express it as a single class one has to choose a surjective homomorphism $\varphi:K_1\times\cdots\times K_n\rightarrow K$ onto a common group $K$ and, with it, the covering group $\mathbb{G}=(\mathbb{G}_1\times\cdots\times\mathbb{G}_n)/\ker\varphi$ of $(\times)_{i=1}^nG_i$ to which the lift is sought. With this choice we get
\begin{eqnarray}
q(\alpha,\beta,\lambda)=\varphi\big(q^1(\alpha,\beta,\lambda),\ldots,q^n(\alpha,\beta,\lambda)\big),
\end{eqnarray} rendering thus
\begin{eqnarray}
\kappa[(\oplus_W)_{i=1}^n P_i]=\varphi_{*}(\kappa^1,\ldots,\kappa^n).\label{1212}
\end{eqnarray} For $K_i=\mathbb{Z}_2$ and $\varphi$ the sum this is $\kappa=\sum_{i}\kappa^i$. Within this context, observe that the choice of $\varphi$ is part of the procedure. There is no canonical choice a priori, but in the geometric situations below, $\varphi$ is dictated by an ambient group in which the product embeds, such as $\GLp{m}\times\GLp{m}\subset\GLp{2m}$ or $SO(m)\times SO(m)\subset SO^{e}_{m,m}$, and by the corresponding covering of that group. In this vein, it can be readily seen that for two oriented vector bundles $E_1,E_2$ of ranks $k_1,k_2\geq2$, respectively, over $\mathcal{M}$, $w_2(E_1\oplus E_2)=w_2(E_1)+w_2(E_2)$. 

Some technical details are in order at this point. First, we identify the obstructions $\kappa_i$ with the second Stiefel-Whitney classes $\omega_2(P_i)$ for each $i=1,\dots,n$. This identification is explicitly carried out in Ref. \cite{GP}. Therefore, Eq. (\ref{1212}) is essentially a Whitney sum formula in degree two. Furthermore, due to the classical isomorphism $\bigoplus_{i=1}^n \check{H}^2(\mathcal{M}, K_i) \simeq \check{H}^2(\mathcal{M}, \bigoplus_{i=1}^n K_i)$, one would be tempted to classify $\kappa[(\oplus_W)_{i=1}^n P_i]$ directly as an ordered sequence operating in $\bigoplus_{i=1}^n K_i$. We note, nevertheless, that the Whitney sum allows for a different perspective: given a surjective homomorphism $\varphi$ as above, one may consider $\kappa[(\oplus_W)_{i=1}^n P_i]$ as taking values in a single periodic group. This latter possibility is indeed what we should keep in mind for the remainder of our argumentation. Before specializing to our application, we note in passing that classical results can be recovered with the presented approach. For instance, taking $P_1$ as the $SO(\dim\mathcal{M})$ frame bundle of $T\mathcal{M}$, with $\rho_1:Spin(\dim\mathcal{M})\rightarrow SO(\dim\mathcal{M})$ as the double cover and $P_2$ as the principal $U(1)$ bundle associated with the complex line bundle with $\rho_2: U(1)\rightarrow U(1)$ the squaring map, we have independent individual obstructions, the second Stiefel-Whitney and first Chern class (mod 2), respectively, whose sum must vanish. (Here $\mathbb{G}=Spin(\dim\mathcal{M})\times_{\mathbb{Z}_2}U(1)$, with $\varphi$ the sum). This recovers the standard $Spin^{c}(\dim\mathcal{M})$ existence criterion \cite{bb}.

We are now in a position to extract information about obstructions related to the so-called hyperbolic Clifford bundles and the corresponding hyperbolic structure, as constructed in Ref. \cite{nc}. The hyperbolic tangent bundle $T\mathcal{M} \oplus_W T^*\mathcal{M}$ is associated with the principal frame bundle $F_H(\mathcal{M}) = \cup_{x \in \mathcal{M}} (F_x\mathcal{M} \oplus_W F^*_x\mathcal{M})$. Assuming $\mathcal{M}$ to be orientable (as is physically desirable), the individual components of the Whitney sum are also orientable and the structure group of $T\mathcal{M}$ reduces to the identity component $\mathrm{GL}^{+}(\dim\mathcal{M},\mathbb{R})$. Moreover, the structure group for each individual component is $\mathrm{GL}^{+}(\dim\mathcal{M}, \mathbb{R})$, with $\mathrm{GL}^{+}(2\dim\mathcal{M}, \mathbb{R})$ serving as the structure group counterpart for $F_H$ (through $\iota$ of (\ref{eq:iota}), or through $j$ of (\ref{eq:j}) when $T\mathcal{M}$ and $T^{*}\mathcal{M}$ are regarded as independent summands). Let $\widetilde{\mathrm{GL}}^{+}(\dim\mathcal{M}, \mathbb{R})$ denote the connected double cover of $\mathrm{GL}^{+}(\dim\mathcal{M}, \mathbb{R})$ and let $\rho :\widetilde{\mathrm{GL}}^{+}(\dim\mathcal{M}, \mathbb{R})\rightarrow \mathrm{GL}^{+}(\dim\mathcal{M}, \mathbb{R})$ be the central homomorphism with $\ker \rho = \mathbb{Z}_2$. Therefore, considering $K=\mathbb{Z}_2$ and $\varphi$ the sum, and restricting the covering of $\mathrm{GL}^{+}(2\dim\mathcal{M},\mathbb{R})$ to $\mathrm{GL}^{+}(\dim\mathcal{M},\mathbb{R})\times\mathrm{GL}^{+}(\dim\mathcal{M},\mathbb{R})$, we have
\begin{equation}
\kappa(F_H)=\kappa(T\mathcal{M})+\kappa(T^*\mathcal{M}). 
\end{equation} 
Nevertheless, as is well known, $T\mathcal{M} \cong T^*\mathcal{M}$ (by means of a Riemannian metric, on any paracompact manifold) and, therefore, $\kappa(F_H) = \kappa(T\mathcal{M}) + \kappa(T\mathcal{M})$. It should be stressed, however, that the hyperbolic bundle possesses a unique connected double covering corresponding to $\mathrm{GL}^{+}(2\dim\mathcal{M}, \mathbb{R})$ \cite{nc}. That is to say, from an algebraic point of view, the Whitney sum gives rise to an inclusion map $i: \mathrm{GL}(\dim\mathcal{M},\mathbb{R}) \times \mathrm{GL}(\dim\mathcal{M},\mathbb{R}) \hookrightarrow \mathrm{GL}(2\dim\mathcal{M},\mathbb{R})$ which, in turn, induces \cite{ao} a fundamental group homomorphism $i_* : \pi_1(\mathrm{GL}(\dim\mathcal{M},\mathbb{R})) \times \pi_1(\mathrm{GL}(\dim\mathcal{M},\mathbb{R})) \rightarrow \pi_1(\mathrm{GL}(2\dim\mathcal{M},\mathbb{R}))$. Since $\pi_1(\mathrm{GL}(k,\mathbb{R})) \cong \mathbb{Z}_2$ for $k \geq 3$ (here and below the groups $\mathrm{GL}$ are understood as their identity components $\mathrm{GL}^{+}$, which are the structure groups once $\mathcal{M}$ is oriented, and $\dim\mathcal{M}\geq3$ is assumed in this argument; the map $i$ is the block-diagonal map $j$ of (\ref{eq:j}), and its effect on $\pi_1$ is computed in Lemma~\ref{lem:pi1}(i)), this leads to
\begin{eqnarray}
i_*&:&\left. \mathbb{Z}_2\times \mathbb{Z}_2\rightarrow \mathbb{Z}_2\right.\nonumber\\&&
\left. (x,y)\mapsto x+y \;\;\; (\!\!\!\!\!\mod 2).\right.
\end{eqnarray} Hence, $\kappa(F_H)$ is an element of $\check{H}^2(\mathcal{M}, \mathbb{Z}_2)$. Thus, $\kappa(F_H)$ takes values in a single $\mathbb{Z}_2$ group, leading to
\begin{equation}
\kappa(F_H)=\kappa(T\mathcal{M})+\kappa(T\mathcal{M})=0 \mod 2,
\end{equation} from which we see that there is no obstruction to the $\widetilde{\mathrm{GL}}(2\dim\mathcal{M},\mathbb{R})$-structure for $F_H$. 

The argument above uses fundamental groups and therefore requires $\dim\mathcal{M}\geq3$. The groups $\widetilde{\mathrm{GL}}$ and $\mathrm{GL}$ should be read as $\GLpt{\cdot}$ and $\GLp{\cdot}$, respectively. The results are collected in the following theorem, whose proof through the Whitney formula depends neither on the dimension nor on $\pi_1$.

\begin{theo}\label{thm:main}
Let $\mathcal{M}$ be an oriented smooth manifold of dimension $m\geq2$, $\HM$ its hyperbolic tangent bundle with the canonical neutral form, and $F_H(\mathcal{M})$ its hyperbolic frame bundle, a principal $\GLp{2m}$-bundle. Then:
\begin{itemize}
\item[(a)] $w_2(\HM)=0$;
\item[(b)] $F_H(\mathcal{M})$ admits a $\GLpt{2m}$-structure, that is, its structure group lifts to the connected double covering of $\GLp{2m}$;
\item[(c)] $\HM$ admits a $Spin^{e}_{m,m}$-structure, that is, its $SO^{e}_{m,m}$-bundle of oriented orthonormal frames lifts to $Spin^{e}_{m,m}$;
\item[(d)] the equivalence classes of the structures in (b), and of those in (c), form a torsor under $\check{H}^1(\mathcal{M},\mathbb{Z}_2)$. In particular they are unique if $\pi_1(\mathcal{M})=0$.
\end{itemize}
No condition on $w_2(\mathcal{M})$ is required.
\end{theo}

Part (c) is proved in Section~\ref{sec:ortho}. For (a) and (b) there are two independent proofs: the one given above, through fundamental groups ($m\geq3$), and the following one, through the Whitney product formula ($m\geq2$).

\begin{proof}[Proof of (a) and (b)]
Choose a Riemannian metric on $\mathcal{M}$. It induces an isomorphism $T\mathcal{M}\simeq T^{*}\mathcal{M}$, so $w(T^{*}\mathcal{M})=w(T\mathcal{M})$ and, by the Whitney product formula \cite{ms},
\[
w(\HM)=w(T\mathcal{M})\,w(T^{*}\mathcal{M})=w^2(T\mathcal{M}).
\]
In characteristic $2$ one has $(1+w_1+w_2+\cdots)^2=1+w_1^2+w_2^2+\cdots$, so that
\begin{equation}
w_1(\HM)=0,\qquad w_2(\HM)=w^2_1(\mathcal{M}) .\label{eq:w2}
\end{equation}
Since $\mathcal{M}$ is oriented, $w_1(\mathcal{M})=0$ and hence $w_2(\HM)=0$, which is (a). The bundle $\HM$ is oriented, of rank $2m\geq4$, and $\GLpt{2m}\rightarrow\GLp{2m}$ restricts to $Spin(2m)\rightarrow SO(2m)$ on maximal compact subgroups. By the criterion recalled in Section~\ref{sec:obstr} and the identification $\kappa=w_2$, $F_H(\mathcal{M})$ admits a $\GLpt{2m}$-structure if, and only if, $w_2(\HM)=0$. This proves (b).
\end{proof}

\begin{proof}[Proof of (d)]
For a central homomorphism with kernel $\mathbb{Z}_2$ over a principal bundle that admits a lift, the equivalence classes of lifts are in bijection with $\check{H}^1(\mathcal{M},\mathbb{Z}_2)$, in a manner compatible with the action of $\check{H}^1(\mathcal{M},\mathbb{Z}_2)$ by changing the lifts $\gamma_{\alpha\beta}\mapsto\gamma_{\alpha\beta}k_{\alpha\beta}$, $k\in Z^1(\mathcal{M},\mathbb{Z}_2)$ \cite{GP,bb}. Hence the set of equivalence classes is a torsor, and for $\pi_1(\mathcal{M})=0$ one has $\check{H}^1(\mathcal{M},\mathbb{Z}_2)=0$.
\end{proof}

Note that the isomorphism $T\mathcal{M}\simeq T^{*}\mathcal{M}$ holds on every paracompact manifold and orientability enters only through $w_1(\mathcal{M})=0$, which kills the cross term in $w_2(T\mathcal{M}\oplus T^{*}\mathcal{M})=w_2(T\mathcal{M})+w_2(T^{*}\mathcal{M})+w_1(T\mathcal{M})w_1(T^{*}\mathcal{M})$, and through the reduction of the structure group of $T\mathcal{M}$ to $\GLp{m}$. The algebraic content of the argument through fundamental groups is that the double covering of $\GLp{2m}$ is trivial when restricted to the subgroup $\iota(\GLp{m})$, because the two blocks $A$ and $(A^{-1})^{T}$ contribute the same element of $\pi_1\simeq\mathbb{Z}_2$ and $x+x=0$. By the lifting criterion for covering spaces, $\iota$ lifts to a homomorphism $\tilde\iota:\GLp{m}\rightarrow\GLpt{2m}$ if, and only if, $\iota_{*}\pi_1(\GLp{m})\subset p_{*}\pi_1(\GLpt{2m})=0$, which holds by Lemma \ref{lem:pi1} (iii). The bundle $F^{+}(\mathcal{M})\times_{\tilde\iota}\GLpt{2m}$ is then a $\GLpt{2m}$-structure on $F_H(\mathcal{M})$, whatever the obstruction class of $F^{+}(\mathcal{M})$.

We shall emphasize once again that the proof through the Whitney formula is independent of the dimension, $m\geq2$ (so that $2m\geq3$). The proof through fundamental groups uses $\pi_1(\GLp{m})\simeq\mathbb{Z}_2$ and therefore requires $m\geq3$. 

We remark that Part (c) of Theorem~\ref{thm:main} is a proposition of \cite{gualth}, which in turn rests on Karoubi's criterion \cite{kar}. Three differences should be noted. First, the orientability required in Karoubi's criterion, as it is reproduced in \cite{gualth}, is that of the bundle $E=T\oplus T^{*}$, which carries a canonical orientation on every manifold. The assumption that $\mathcal{M}$ be oriented, which we make, is stronger than what the $Spin(m,m)$ statement requires. It is needed for the $\GLp{2m}$-covering of parts (a), (b). Second, part (b) concerns the covering of the full linear group, which is the setting of the hyperbolic frame bundle of \cite{nc} and is not discussed in the generalized geometry literature. Third, our proofs of (a)--(c) use only the Whitney product formula and the homomorphism $\iota$ on $\pi_1$, and not the criterion for indefinite signature. The splitting of the spin covering over $\mathrm{GL}(V)$ described in \cite{gualth} is the $Spin^{e}_{m,m}$ counterpart of the lift $\tilde\iota$ above (see Section~\ref{sec:ortho}).

An important aspect of the hyperbolic construction is the existence of a metric field, by means of which orthonormal frames may be introduced. This procedure leads to a remarkable reduction of the frame bundle to, among other possibilities, the (special and oriented) hyperbolic frame bundle, whose structure group is $SO^e_{\mathbb{R}^{p,q}\oplus(\mathbb{R}^*)^{p,q}}$.
The canonical neutral form has signature $(m,m)$ whatever the signature $(p,q)$ of any auxiliary metric on $T\mathcal{M}$, so this group must preserve isometries in the hyperbolic space \cite{nc} and, hence, it is $SO^{e}_{m,m}$. The passage from the general linear structure group to the orthonormal one is not automatic: the inverse image of the reduced group under the covering has to be examined. This is done in Section~\ref{sec:ortho}, where the absence of obstructions to a $Spin^{e}_{m,m}$ structure is proved (Theorem~\ref{thm:main}(c)), whose relation with the results of \cite{kar,gualth,gualt} was mentioned above. At this point, it is instructive to place these results in a broader context. The comprehensive study in the realm of complex geometry \cite{gualt} addresses structure group reductions $O(2n,2n) \to U(n,n)$ and integrability, in which the spinor bundle $S=\wedge^\bullet T^*\mathcal{M}\otimes(\det T)^{1/2}$ plays a central role. The global existence of this bundle on every manifold is established in \cite{gualth}. It is recovered here in Section~\ref{sec:ortho} from the $Spin^{e}_{m,m}$ structure of Theorem~\ref{thm:main} (c).

Finally, we remark that standard results regarding uniqueness apply: the spinor structure is unique if $\pi_1(\mathcal{M})=0$, but when the base manifold possesses a non-trivial fundamental group, there will be as many non-equivalent spinor structures as there are distinct elements in $\check{H}^1(\mathcal{M},\mathbb{Z}_2)$. More precisely, the equivalence classes form a torsor under $\check{H}^1(\mathcal{M},\mathbb{Z}_2)$ (Theorem~\ref{thm:main}(d)).

\section{\texorpdfstring{Orthonormal frames and $Spin^{e}_{m,m}$}{Orthonormal frames and Spin(m,m)}}\label{sec:ortho}

We now delve into the question of reduction to the maximal compact subgroup. Staring from $\mathcal{M}$ oriented, since $\iota(\GLp{m})\subset SO^{e}_{m,m}$, the bundle of oriented $\langle\,,\rangle$-orthonormal frames of $\HM$ contains the principal $SO^{e}_{m,m}$-subbundle
\begin{equation}
P=F^{+}(\mathcal{M})\times_{\iota}SO^{e}_{m,m}.\label{eq:P}
\end{equation}
The canonical neutral form has signature $(m,m)$ whatever the signature of any auxiliary metric on $\mathcal{M}$. For this reason the groups appearing here are $SO^{e}_{m,m}$ and $Spin^{e}_{m,m}$, and not groups of the type $SO^{e}_{p,q}$ with $p+q=2m$ arbitrary. Fix a Riemannian metric $g$ on $\mathcal{M}$ and define $E^{\pm}\subset\HM$ fibrewise as in (\ref{eq:Epm}). The projection onto $T\mathcal{M}$ restricts to isometries $(E^{\pm},\pm\frac12\langle\,,\rangle)\simeq(T\mathcal{M},g)$, which orient $E^{\pm}$. Frames of $P$ whose first $m$ vectors span $E^{+}$ and whose last $m$ span $E^{-}$ (so-called adapted frames) form a principal $SO(m)\times SO(m)$-subbundle $Q\subset P$, a reduction of $P$ to the maximal compact subgroup. Since $E^{\pm}$ are both isometric to $(T\mathcal{M},g)$, we have
\begin{equation}
Q\simeq Q_0\times_{\mathcal{M}}Q_0=\{(p, p') \in Q_0 \times Q_0 | \pi(p) = \pi(p')\},\label{eq:Q}
\end{equation}
where $Q_0$ is the principal $SO(m)$-bundle of oriented $g$-orthonormal frames of $T\mathcal{M}$. We are now in position to prove two results. 

\begin{lemma}\label{lem:maxcompact}
\begin{itemize}
\item[(i)] The preimage of $SO(m)\times SO(m)$ under $\rho:Spin^{e}_{m,m}\rightarrow SO^{e}_{m,m}$ is $Spin(m)\times_{\mathbb{Z}_2}Spin(m)=(Spin(m)\times Spin(m))/\{(1,1),(-1,-1)\}$.
\item[(ii)] The bundle $P$ admits a $Spin^{e}_{m,m}$-structure if, and only if, its reduction $Q$ admits a $Spin(m)\times_{\mathbb{Z}_2}Spin(m)$-structure.
\end{itemize}
\end{lemma}

\begin{proof}
(i) Using $E^{+}\perp E^{-}$, the even elements of $Cl(E^{+})$ and $Cl(E^{-})$ commute. Take $x\in Spin(E^{+})$, the group of even products of vectors of $E^{+}$ of unit norm, and $y\in Spin(E^{-})$, the even products of vectors of $E^{-}$ of norm $-1$ (for the Clifford relation of the algebra considered); each group is isomorphic to $Spin(m)$. The map $(x,y)\mapsto xy$ is a homomorphism into $Spin^{e}_{m,m}$ covering $(\rho(x),\rho(y))\in SO(E^{+})\times SO(E^{-})$. If $xy=1$ then $y=x^{-1}\in Cl^{+}(E^{+})\cap Cl^{+}(E^{-})=\mathbb{R}\cdot1$, so $x=\pm1$ and $y=x^{-1}=\pm1$ with the same sign. The kernel of the map is therefore $\{(1,1),(-1,-1)\}$ and its image is the preimage of $SO(m)\times SO(m)$. (ii) Choose an open cover $\{U_\alpha\}$ of $\mathcal{M}$ with local sections $\sigma_\alpha$ of $Q$. They are also local sections of $P$, so $Q$ and $P$ have the same transition functions, say $s_{\alpha\beta}$, with values in $H=SO(m)\times SO(m)\subset SO^{e}_{m,m}$. By the criterion recalled in Section~\ref{sec:obstr}, $P$ admits a $Spin^{e}_{m,m}$-structure if, and only if, there are continuous maps $\gamma_{\alpha\beta}:U_\alpha\cap U_\beta\rightarrow
Spin^{e}_{m,m}$ with $\rho(\gamma_{\alpha\beta})=s_{\alpha\beta}$ satisfying the same consistency relations as the $s_{\alpha\beta}$ and $Q$ admits a $Spin(m)\times_{\mathbb{Z}_2}Spin(m)$-structure
if, and only if, the same holds with $\gamma_{\alpha\beta}$ taking values in $\rho^{-1}(H)=Spin(m)\times_{\mathbb{Z}_2}Spin(m)$, by (i). Since $\rho(\gamma_{\alpha\beta})=s_{\alpha\beta}\in H$
forces $\gamma_{\alpha\beta}\in\rho^{-1}(H)$, the two conditions coincide.
\end{proof}

\begin{prop}\label{prop:spin0}
Let $\mathcal{M}$ be oriented, $m\geq2$. The bundle $P$ of (\ref{eq:P}), and hence $\HM$, admits a $Spin^{e}_{m,m}$-structure. This is part (c) of Theorem~\ref{thm:main}.
\end{prop}

\begin{proof}
By Lemma~\ref{lem:maxcompact} (ii) it suffices to lift $Q\simeq Q_0\times_{\mathcal{M}}Q_0$ to $Spin(m)\times_{\mathbb{Z}_2}Spin(m)$. Taking $K_1=K_2=\mathbb{Z}_2$ and $\varphi$ the sum, so that  $\mathbb{G}=Spin(m)\times_{\mathbb{Z}_2}Spin(m)$, the obstruction is $\kappa^1+\kappa^2$, where $\kappa^1=\kappa^2=\kappa(Q_0)=w_2(\mathcal{M})$ because the two factors of $Q$ are \emph{the same} bundle $Q_0$. Hence the obstruction is $2w_2(\mathcal{M})=0$.
\end{proof}

We stress that in the notation of Karoubi's criterion \cite{kar,gualth}, the obstruction found here is $w_2(E^{+})+w_2(E^{-})$, and the criterion $w_2(E^{+})=w_2(E^{-})$ is recovered. In the present case it is a direct consequence of the construction with no further input than the identification $\kappa=w_2$ for $Spin(m)\rightarrow SO(m)$. This is what we mean by saying that the known result is obtained by an elementary route.

Going further, Gualtieri \cite{gualth} gives an explicit homomorphism $\hat\iota:\GLp{m}\rightarrow Spin^{e}_{m,m}$ with $\rho\circ\hat\iota=\iota$, a splitting of the spin covering over $\mathrm{GL}^{+}(V)$ obtained by a Clifford computation. We write
$\hat{P}=F^{+}(\mathcal{M})\times_{\hat\iota}Spin^{e}_{m,m}$ for the associated $Spin^{e}_{m,m}$-structure on $\HM$. The Clifford algebra $Cl(\HM)$ acts on $\bigwedge^{\bullet}T^{*}\mathcal{M}$ by $(x_{*}+x^{*})\cdot\varphi=i_{x_{*}}\varphi+x^{*}\wedge\varphi$, and the spinor representation of $Spin^{e}_{m,m}\subset Cl(H_V)$ on $\bigwedge^{\bullet}V^{*}$ decomposes into even and odd forms. Restricted to $\hat\iota(\GLp{m})$ it is given by \cite{gualth}
\begin{equation}
g\cdot\varphi=\sqrt{\det g}\,\big(g^{*}\big)^{-1}\varphi,\qquad g\in\GLp{m},\label{eq:spinrep}
\end{equation}
with the positive square root. This is a genuine representation precisely because the covering is trivial over $\iota(\GLp{m})$. The associated spinor bundle is therefore
\begin{equation}
S=\hat{P}\times_{Spin^{e}_{m,m}}\bigwedge\nolimits^{\bullet}\mathbb{R}^{m*}\;\simeq\;\bigwedge\nolimits^{\bullet}T^{*}\mathcal{M}\otimes(\det T\mathcal{M})^{1/2},\label{eq:S}
\end{equation}
where $(\det T\mathcal{M})^{1/2}$ is the real line bundle associated with $g\mapsto(\det g)^{1/2}>0$. It is trivial, because its structure group is the contractible group $(0,\infty)$, so that $S\simeq\bigwedge^{\bullet}T^{*}\mathcal{M}$ as a vector bundle. Moreover, $Spin^{e}_{m,m}$ lies in the even subalgebra of $Cl(H_V)$ and each vector changes the degree of a form by $\pm1$, so the representation preserves the parity of forms. Hence $S=S^{+}\oplus S^{-}$, with $S^{+}\simeq\bigwedge^{\mathrm{even}}T^{*}\mathcal{M}$
and $S^{-}\simeq\bigwedge^{\mathrm{odd}}T^{*}\mathcal{M}$. The existence of $S$ requires no condition beyond those of Theorem~\ref{thm:main}. In particular, no condition on $w_2(\mathcal{M})$. It is important to emphasize that the bundle $S$ is of rank $2^m$ and consists of differential forms. It is not the Dirac spinor bundle of $(\mathcal{M},g)$, which has rank $2^{\lfloor m/2\rfloor}$ (for $m$ even, $\mathbb{C}$-rank $2^{m/2}$) and exists if, and only if, $w_2(\mathcal{M})=0$. A $Spin^{e}_{m,m}$-structure on $\HM$ carries no information on $w_2(T\mathcal{M})$, and the existence of $S$ does not remove the obstruction to an ordinary spin structure on $T\mathcal{M}$. If $\mathcal{M}$ is spin and $m$ is even, then $\bigwedge^{\bullet}T^{*}\mathcal{M}\otimes\mathbb{C}\simeq Cl(T\mathcal{M})\otimes\mathbb{C}\simeq S_D\otimes S_D^{*}$ as bundles, where $S_D$ is the Dirac spinor bundle \cite{bb}. This is nothing but the classical identification of Dirac--K\"ahler fields with bispinors \cite{kah,bj}. The point is that the left-hand side exists without the spin hypothesis.

We shall finalize this section by making some remarks on the physical interpretation. The hyperbolic Clifford algebra was introduced to treat multivectors and multiforms on the same footing, and for the identification of $\bigwedge V^{*}$ with a minimal left ideal whose elements represent Witten superfields \cite{RSV,RS}. The results above show that the bundle structure behind this construction is globally well defined on every oriented manifold, and that the splitting $E^{+}\oplus E^{-}$ realizes the standard Clifford bundles $Cl(T\mathcal{M},\pm g)$ of the Clifford bundle formalism \cite{RS0,NR} inside $Cl(\HM)$. No claim is made, however, about spinor fields of Dirac type on manifolds that do not admit a spin structure, nor about a replacement for a $Spin^{c}$ structure or for a twisting gauge field: $S$ is not the bundle on which the Dirac equation is posed. For a more physically oriented reader, the spinors of $Spin(d,d)$ on $T\oplus T^{*}$ are the ones that appear in generalized geometry and in the formulation of supergravity in terms of generalized geometry, where differential forms play the role of spinors \cite{gualt,csw}.

\section{\texorpdfstring{Conclusions}{Conclusions}}\label{sec:concl}

The results presented in this paper demonstrate that regardless of the individual obstructions to the lifting of $T\mathcal{M}$ and $T^*\mathcal{M}$, they compensate for each other. Consequently, the second Stiefel-Whitney class associated with their Whitney sum vanishes, enabling (following a frame bundle reduction) the existence of spinor structures in hyperbolic Clifford bundles without any topological obstruction provided $\mathcal{M}$ is orientable and $\dim\mathcal{M}\geq 2$ (the argument through fundamental groups requires $\dim\mathcal{M}\geq3$, while the one through the Whitney formula does not). Specifically, the orientability of $\mathcal{M}$ was used in Theorem~\ref{thm:main} to have $w_1(\mathcal{M})=0$. We now show that this hypothesis cannot be removed for the covering of $\GLp{2m}$.
\begin{prop}\label{prop:nonor}
Let $\mathcal{M}$ be a smooth manifold of dimension $m\geq2$, not necessarily orientable. The bundle $\HM$ is orientable, and its oriented frame bundle admits a $\GLpt{2m}$-structure if, and only if, $w_1^2(\mathcal{M})=0$.
\end{prop}
\begin{proof}
By (\ref{eq:w2}), which was derived without using the orientability of $\mathcal{M}$, $w_1(\HM)=0$ and $w_2(\HM)=w_1^2(\mathcal{M})$. The statement follows from the criterion recalled in Section~\ref{sec:obstr} and the identification $\kappa=w_2$.
\end{proof}

The obtained conclusions are known \cite{kar,gualth} and here we have shown that they follow by elementary means, namely, the Whitney product formula and the isomorphism $T\mathcal{M}\simeq T^{*}\mathcal{M}$ and the fact that $A\mapsto(A^{-1})^T$ induces the identity on $\pi_1(\mathrm{GL}^{+}(\dim\mathcal{M},\mathbb{R}))\simeq\mathbb{Z}_2$, so that the double covering of $\mathrm{GL}^{+}(2\dim\mathcal{M},\mathbb{R})$ is trivial over the image of $\mathrm{GL}^{+}(\dim\mathcal{M},\mathbb{R})$. The $Spin^{e}_{m,m}$ structure follows from the construction because the reduction of the hyperbolic frame bundle to the maximal compact subgroup is the fibre product of two copies of the same bundle of frames, with no need of Karoubi's criterion for indefinite signature. We believe this simplification is of some pedagogical and conceptual interest and it answers the question left open in \cite{nc}. We have also clarified that the spinor bundles of $T\mathcal{M}\oplus T^{*}\mathcal{M}$ are bundles of differential forms, not Dirac spinor bundles, that their existence says nothing about $w_2(\mathcal{M})$, and that orientability matters for the covering of the linear group but not for $Spin(m,m)$. The geometrical setup of hyperbolic Clifford bundles is, therefore, of algebraic interest for the study of supersymmetry, superfields, and (topological) quantum and string field theories, as motivation only.

The main natural ensuing question now is the construction of spinor fields and Dirac operators associated with the hyperbolic Clifford algebra. These aspects are of prime importance and we shall approach them in future works.

\section*{Acknowledgments}

JMHS thanks CNPq (grant No. 307641/2022-8) for financial support.

\section*{Declaration of generative AI and AI-assisted technologies in the manuscript preparation process}

During the preparation of this work the authors used Gemini (Google) as a language reviewer. In the new version of the manuscript we used Claude (Anthropic) to help framing the discussion in terms of frame bundle reduction and organize the responses to Referees. After using this tool/service, the authors reviewed and edited the content as needed and take full responsibility for the content of the article.

\end{document}